\documentclass{article}

\usepackage{amsmath,amssymb,amsfonts}   
\usepackage{graphicx,bm}

\newcommand{\calP}{{\mathcal P}}

\newcommand{\calM}{{\mathcal M}}

\newcommand{\R}{{\mathbb R}}

\renewcommand{\P}{\mathbb{P}}
\newcommand{\PP}{\widetilde{P}}
\newcommand{\Q}{\widetilde{Q}}
\newcommand{\x}{\mathbf{x}}
\newcommand{\y}{\mathbf{y}}
\newcommand{\e}{{\mathrm e}}

\newcommand{\n}{\mathbf n}
\newcommand{\calT}{{\mathcal T}}
\renewcommand{\P}{\mathbb P}
\newcommand{\p}{\widetilde{p}}

\begin{document}

 \title{A probabilistic model of diffusion through a semi-permeable barrier}

\author{ \em
P. C. Bressloff, \\ Department of Mathematics, 
University of Utah \\155 South 1400 East, Salt Lake City, UT 84112}

 \maketitle

\begin{abstract} 
Diffusion through semipermeable structures arises in a wide range of processes in the physical and life sciences. Examples at the microscopic level range from artificial membranes for reverse osmosis to lipid bilayers regulating molecular transport in biological cells to chemical and electrical gap junctions. There are also macroscopic analogs such as animal migration in heterogeneous landscapes. It has recently been shown that one-dimensional diffusion through a barrier with constant permeability $\kappa_0$ is equivalent to snapping out Brownian motion (BM). The latter sews together successive rounds of partially reflecting BMs that are restricted to either the left or right of the barrier. Each round is killed when its Brownian local time exceeds an exponential random variable parameterized by $\kappa_0$. A new round is then immediately started in either direction with equal probability. In this paper we use a combination of renewal theory, Laplace transforms and Green's function methods to show how an extended version of snapping out BM provides a general probabilistic framework for modeling diffusion through a semipermeable barrier. This includes modifications of the diffusion process away from the barrier (eg. stochastic resetting) and non-Markovian models of membrane absorption that kill each round of partially reflected BM. The latter leads to time-dependent permeabilities.\end{abstract}

\section{Introduction}

Diffusion through semipermeable barriers or membranes arises in a wide range of processes in the physical and life sciences. At the microscopic level, 
a semipermeable membrane is a biological or artificial membrane that only allows certain molecules to pass through it. This can be quantified more precisely in terms of the membrane permeability, which is the passive diffusion rate of molecules across the membrane. The permeability of any specific molecule depends on properties such as its size and ionic charge. Artificial semipermeable membranes include a variety of materials that are specifically designed for filtration. A well known example is water filtration via reverse osmosis.  There are many examples of permeable structures in biological cells, which regulate the flow of proteins and ions between different subcellular compartments and the exchange of molecules with the extracellular environment \cite{Philips12,Alberts15,Bressloff21}. Molecular transport is typically mediated by protein-based pores embedded in the lipid bilayer of the plasma membrane and membrane-bound organelles. In addition, scaffolding proteins within the plasma membrane act as semi-permeable barriers to lateral diffusion  \cite{Kusumi05}. An important example of a semi-permeable barrier at the multicellular level is a gap junction. Gap junctions are small nonselective channels that provide a direct diffusion pathway between neighboring cells. They are formed by the head-to-head connection of two hemichannels or connexons, one from each of the two coupled cells  \cite{Evans02,Connors04,Good09}. Gap junctions are prevalent in most animal organs and tissues, providing a mechanism for both electrical and chemical communication between cells. Finally, permeable barriers are found at the ecological level where, for example, animal dispersal is affected by the presence of roads and fences within a heterogeneous landscape \cite{Beyer16,Assis19,Kenkre21}.

The classical boundary condition for a semi-permeable membrane takes the flux across the membrane to be continuous and to be proportional to the difference in concentrations on either side of the barrier \cite{Tanner78,Brink85,Ramanan90,Kos01,Moutal19}; the constant of proportionality is the permeability. For example, consider one-dimensional diffusion with a semipermeable barrier at $x=0$. Let $u(x,t)$ be the concentration at position $x\in \R$ at time $t$. The boundary value problem (BVP) takes the form
\begin{subequations}
\label{class}
\begin{align}
\frac{\partial u(x,t)}{\partial t}&=D\frac{\partial^2 u(x,t)}{\partial x^2}, \quad x\neq 0,\\
J(0^{\pm},t)&=\kappa_0[u(0^-,t)-u(0^+,t)],
\end{align}
\end{subequations}
where $J(x,t)=-D\partial_xu(x,t)$, $D$ is the diffusivity and $\kappa_0$ is the (constant) permeability. Equation (\ref{class}b) is known as the permeable or leather boundary condition, and $u(x,t)$ is understood as a weak solution. One limitation of this macroscopic model is that it is not based on a fundamental microscopic theory of single-particle diffusion. This has motivated a number of studies of random walks on lattices in which semipermeable barriers are represented by local defects \cite{Powles92,Kenkre08}. Moreover, a Fokker-Planck description of single-particle diffusion through a semiperrmeable membrane has recently been derived by taking an appropriate continuum limit of a random walk model \cite{Kay22}. An alternative approach to modeling single-particle diffusion is to use stochastic differential equations (SDEs). It has been known for a long time that in order to formulate Brownian motion (BM) in a bounded domain, it is necessary to modify the standard Wiener process. For example, one can implement totally and partially reflecting boundaries by introducing a Brownian functional known as the boundary local time \cite{Ito65,Freidlin85,Papanicolaou90,Milshtein95,Borodin96,Grebenkov06}. The latter determines the amount of time that a Brownian particle spends in the neighborhood of points on the boundary. (In terms of the Fokker-Planck description, a totally (partially) reflecting boundary corresponds to a Neumann (Robin) boundary condition.) The extension of one-dimensional BM to include a semipermeable barrier is more recent, and is based on so-called snapping out BM \cite{Lejay16}. Snapping out BM involves sewing together two partially reflecting BMs, one restricted to $x<0$ and the other restricted to $x>0$. Suppose that the particle starts in the domain $x>0$. It realizes positively reflected BM until its local time exceeds an exponential random variable with parameter $\kappa_0$. It then immediately resumes either negatively or positively reflected BM with equal probability, and so on. Note that snapping out BM is related to the more familiar skew BM first introduced by Ito and McKean \cite{Ito63}. Skew BM evolves as
standard BM reflected at the origin so that the next excursion is chosen
to be positive with a fixed probability $p$. It has a wide range of applications, particularly in mathematical finance \cite{Lejay06,Decamps06,App11,Gairat17}.

In this paper we show how the snapping out BM introduced in Ref. \cite{Lejay16} can be used to develop more general probabilistic models of one-dimensional diffusion through semi-permeable membranes. (Possible extensions to higher spatial dimensions are discussed in section 5.) We begin, in section 2, by describing how to formulate totally and partially reflecting BM in terms of Brownian local times. We then derive a last renewal equation that relates the probability density of snapping out BM with the corresponding probability density for partially reflected BM. The renewal equation is solved using Laplace transforms and Green's function methods, resulting in an explicit expression for the probability density of snapping out BM. We thus establish  that the probability density satisfies equation (\ref{class}). Note that our renewal method is equivalent to the resolvent operator formulation of Ref. \cite{Lejay16}, since they both rely on the strong Markov property. However, expressing the dynamics in terms of a renewal process facilitates the various extensions considered in the remainder of the paper. In section 3 we extend the snapping out BM by incorporating the effects of stochastic resetting, whereby the position of the particle is randomly reset according to a Poisson process with resetting rate $r$. Stochastic resetting has become an important paradigm for understanding nonequilibrium stochastic processes, with a variety of applications in optimal search problems and biophysics (see the review \cite{Evans20} and references therein.) One of the particularly useful features of stochastic resetting is that it can be applied to virtually any stochastic process. In addition, if resetting erases all previous history of particle position then renewal theory can be used to obtain explicit analytical solutions. As far as we are aware, the problem of diffusion through a semipermeable membrane with resetting has not been considered before. One nontrivial feature of this example is that there are two distinct renewal processes, one associated with position resetting and the other with each round of absorption and restart at the membrane interface. 
We show how to modify the renewal equation of snapping out BM and use this to calculate the nonequilibrium stationary state (NESS) in the presence of resetting. We show that the NESS is independent of $\kappa_0$, but that relaxation to the NESS is $\kappa_0$-dependent. In section 4, we combine snapping out BM with the so-called encounter-based model of partial absorption \cite{Grebenkov20,Grebenkov22,Bressloff22,Bressloff22a}. The basic idea is to to kill a given round of partially reflecting BM when the local time exceeds a non-exponential rather than an exponential random variable. In order to determine the probability density we construct a first rather than a last renewal equation. We show that the corresponding boundary condition at the interface involves a time-dependent permeability with memory.  Finally, in section 5 we indicate how to extend the theory to higher spatial dimensions.

\setcounter{equation}{0}
\section{The snapping out Brownian motion}

In order to develop a general probabilistic model of a semi-permeable membrane, we first need to consider the probabilistic version of the one-dimensional model (\ref{class}) based on the snapping out BM introduced by Lejay \cite{Lejay16}. One of the key ingredients is formulating partially reflecting BM on $[0,\infty)$ in terms of the local time $L_t$ at $x=0$.

\subsection{Partially reflected Brownian motion} Let $W$ be a Wiener process on $\R$ and define totally reflected BM according to the function $X_t=F(W_t)\equiv \sqrt{2D}|W_t|$. In order to determine the stochastic differential equation (SDE) for $X_t$, we use the standard Ito formula \cite{Ito65,Freidlin85,Borodin96}
\begin{equation}
\label{Ito}
dX_t=f'(W_t)dW_t+\frac{1}{2}f''(W_t)dt.
\end{equation}
These derivatives are understood in the distributional sense. That is,
\[f'(x)=\sqrt{2D}\, \mbox{sgn}(x),\quad f''(x)=2\sqrt{2D}\delta(x) ,\]
where $\mbox{sgn}(x)= -1 $ for $x\leq 0$ and $+1$ for $ x>0$.
Hence
\begin{align}
dX_t&
=\sqrt{2D}\mbox{sgn}(W_t)dW_t+D\delta(X_t)dt ,
\end{align}
with $\delta(X_t)$ defined on the half-line.
Integrating with respect to time implies that
\[X_t = \int_0^t\mbox{sgn}(W_s)dW_s+L_t,\]
where
\begin{equation}
L_t=D\int_0^t\delta(X_s)ds.
\end{equation}
and $dL_t=D\delta(X_t)dt$.
This is the distribution-based version of the local time of $X$ at $x=0$, which is defined as
\begin{equation*}
L_t=\lim_{\epsilon\rightarrow 0^+}\frac{D}{\epsilon}\int_0^tI\{0 \leq X_s\leq \epsilon\} ds
\end{equation*}
where $I$ is the indicator function. It can be shown that $L_t$ exists and is a nondecreasing, continuous function of $t$. Moreover, the corresponding probability density $p(x,t|x_0)$, $p(x,t|x_0)dx=\P[x\leq X_t < x+dx|\, X_0=x_0]$, satisfies the diffusion equation on $[0,\infty)$ with the totally reflecting boundary condition $J(0,t)=0$. Here $J(x,t)=-D\partial_xp(x,t)$ is the probability flux.

Partially reflected BM, also known as elastic BM, combines reflected BM $X_t$ with a stopping condition that halts the stochastic process when the local time $L_t(X)$ exceeds a random exponentially distributed threshold $\widehat{\ell}$ \cite{Grebenkov06}. That is, the particle is absorbed at $x=0$ at the stopping time
 \begin{equation}
\label{exp}
{\mathcal T}=\inf\{t>0:\ L_t >\widehat{\ell}\},\quad \P[\widehat{\ell}>\ell]\equiv \Psi(\ell) =\e^{-\kappa_0 \ell/D}.
\end{equation}
It can then be shown that the marginal density for particle position (prior to absorption), 
\[p(x,t|x_0)dx=\P[x\leq X_t < x+dx ,\ t < {\mathcal T}|X_0=x_0],\]
satisfies the diffusion equation with a Robin boundary condition at $x=0$ \cite{Grebenkov06}:
 \begin{subequations}
\label{Robin}
\begin{align}
\frac{\partial p(x,t|x_0)}{\partial t}&=D\frac{\partial^2 p(x,t|x_0)}{\partial x^2}, \quad x> 0,\\
D\partial_xp(0,t|x_0)&=\kappa_0 p(0,t|x_0),\quad p(x,0|x_0)=\delta(x-x_0).
\end{align}
\end{subequations}

  \subsection{Brownian motion in the presence of a semipermeable membrane} We now turn to the snapping out BM introduced by Lejay \cite{Lejay16}, and show how it is equivalent to single-particle diffusion through a semipermeable barrier. We proceed by constructing a last renewal equation that relates the probability density of snapping out BM with the corresponding probability density of partially reflected BM. Our analysis is equivalent to the resolvent operator formalism presented in Ref. \cite{Lejay16}, but is more amenable to the generalizations developed in subsequent sections.  

The behavior of the stochastic process is described as follows. Without loss of generality, assume that the particle starts at $X_0=x_0\geq 0$. It realizes positively reflected BM until its local time $L_t$ at $x=0^+$ is greater than an independent exponential random variable $\widehat{\ell}$ of parameter $\kappa_0$. Let $\calT_0$ denote the corresponding stopping time. The process immediately restarts as a new reflected BM with probability 1/2 in either $[0^+,\infty)$ or $(-\infty,0^-]$ and a new local time $\ell_{t_1}$at $x=0^{\pm}$ for $t_1=t-{\mathcal T}_0$. Again the reflected BM is stopped when $\ell_{t_1}$ exceeds a new exponential random variable at the stopping time $\calT_2$ etc. It can be proven that the snapping out BM is a strong Markov process\footnote{Recall that a continuous stochastic process $\{X_t\, \ t\geq 0\}$ is said to have the Markov property if the conditional probability distribution of future states of the process (conditional on both past and present states) depends only upon the present state, not on the sequence of events that preceded it. That is, for all $t'>t$ we have $\P[X_{t'}\leq x|X_{s},s\leq t]=\P[X_{t'}\leq x|X_{t}]$.
 The {strong Markov property} is similar to the Markov property, except that the ``present'' is defined in terms of a stopping time.} on the disjoint space ${\mathbb G}=(-\infty,0^-]\cup [0^+,\infty)$. The strong Markov property means that we can use renewal theory to analyze the evolution of the associated probability density and show that it satisfies the classical semipermeable boundary condition (\ref{class}b). 
 
Let $\rho(x,t|x_0)$ denote the probability density of the snapping out BM with the initial condition $X_0=x_0$ and set 
\begin{equation}
\rho(x,t)=\int_{-\infty}^{\infty}\rho(x,t|x_0)g(x_0)dx_0
\end{equation} 
for any continuous function $g$ on ${\mathbb G}$ with $\int_{-\infty}^{\infty}g(x_0)dx_0=1$. Similarly, set
\begin{equation}
p(x,t)=H(x)\int_{0}^{\infty}p(x,t|x_0)g(x_0)dx_0+H(-x)\int_{-\infty}^{0}p(-x,t|-x_0)g(x_0)dx_0,
\end{equation}
where $H(x)$ is the Heaviside function and $p(x,t|x_0)$ for $x,x_0\geq 0$ is the solution to the Robin BVP (\ref{Robin}). It follows that $\rho(x,0)=p(x,0)=g(x)$. In the special case that $g(x)$ is an even function of $x$, then $\rho(x,t)=\rho(-x,t)$ for all $x\geq 0$ and there is no net flux through the membrane, although individual particles cross the membrane. On the other hand  if $g(x_0)=0$ for $x_0<0$ and $\kappa_0>0$ then $\rho(x,t)$ will have positive definite measure on $(-\infty,0]$ even though $p(x,t)=0 $ for $x<0$ and all $t\geq 0$. (An analogous result holds if $g(x_0)$ vanishes on $[0,\infty)$.)

Given these definitions and the strong Markov property, there exists a last renewal equation of the form
 \begin{align}
  \label{renewal}
 \rho(x,t)&=p(x,t)+\frac{\kappa_0}{2}\int_0^t p(|x|,\tau|0)[\rho(0^+,t-\tau ) +\rho(0^-,t-\tau )]d\tau ,\ x\in {\mathbb G}, \kappa_0 >0.
   \end{align}
 The first term on the right-hand side represents all sample trajectories that have never been absorbed by the barrier at $x=0^{\pm}$ up to time $t$. The corresponding integrand represents all trajectories that were last absorbed (stopped) at time $t-\tau$ in either the positively or negatively reflected BM state and then switched to the appropriate sign to reach $x$ with probability 1/2. Since the particle is not absorbed over the interval $(t-\tau,t]$, the probability of reaching $x \in {\mathbb G}$ starting at $x=0^{\pm}$ is $p(|x|,\tau|0)$. The probability that the last stopping event occurred in the interval $(t-\tau,t-\tau+d\tau)$ irrespective of previous events is $\kappa_0 d\tau$. It is convenient to Laplace transform the renewal equation (\ref{renewal}) with respect to time $t$ by setting $\widetilde{\rho}(x,s) =\int_0^{\infty}\e^{-st}\rho(x,t)dt$ etc. This gives
 \begin{align}
  \label{renewal2}
 \widetilde{\rho}(x,s) = \p(x,s)+\frac{\kappa_0}{2} \p(|x|,s|0)[\widetilde{\rho}(0^+,s )+\widetilde{\rho}(0^-,s ) ],\ x\in {\mathbb G}.
 \end{align}
 (Note that equation (\ref{renewal2}) is equivalent to the resolvent operator equation (8) of \cite{Lejay16}.)
 Setting $x=0^{\pm}$ in equation (\ref{renewal2}), summing the results and rearranging shows that
  \begin{equation*}
 \widetilde{\rho}(0^+,s)+ \widetilde{\rho}(0^-,s) =\frac{ \Gamma(s)}{1-\kappa_0\p(0,s|0)} 
 \end{equation*}
 with $\Gamma(s)\equiv\p(0^+,s)+\p(0^-,s)$.
  Substituting back into equations (\ref{renewal2}) yields the explicit solution
 \begin{align}
  \label{renewal3}
 \widetilde{\rho}(x,s) = \p(x,s)+ \frac{ \kappa_0\Gamma(s)/2}{1-\kappa_0\p(0,s|0)} \p(|x|,s|0) ,\ x\in {\mathbb G}.
 \end{align}
 
 The next step is to evaluate $\p(|x|,s|x_0)$. Laplace transforming equations (\ref{Robin}) shows that $\p(x,s|x_0)$, $x >0$, satisfies the BVP
  \begin{subequations}
\label{RobinLT}
\begin{align}
D\frac{\partial^2 \p(x,s|x_0)}{\partial x^2}-s\p(x,s|x_0)&=-\delta(x-x_0), \quad x> 0,\\
D\frac{\partial \p(0,s|x_0)}{\partial x}&=\kappa_0 \p(0,s|x_0).
\end{align}
\end{subequations}
That is, we can identify $\p(x,s|x_0)$ with the Robin Green's function for the modified Helmholtz equation on $[0,\infty)$. Writing the general solution for $x<x_0$ as
\begin{equation}
\p(x,s|x_0)=A\e^{-\sqrt{s/D}x}+B\e^{\sqrt{s/D}x}
\end{equation}
and substituting into the Robin boundary condition shows that 
\begin{equation}
\p(x,s|x_0)=B\left (\e^{\sqrt{s/D}x}+\frac{\sqrt{sD}-\kappa_0}{\sqrt{sD}+\kappa_0}\e^{-\sqrt{s/D}x}\right ).
\end{equation}
Using the fact that the bounded solution for $x>x_0$ is proportional to $\e^{-\sqrt{s/D}x}$, imposing continuity of $\p(x,s|x_0)$ across $x_0$ and matching the discontinuity in the first derivative yields the solution
\begin{equation}
\label{EBM}
\p(x,s|x_0)=\frac{1}{2\sqrt{sD}}\left (\e^{-\sqrt{s/D}|x-x_0|}+\frac{\sqrt{sD}-\kappa_0}{\sqrt{sD}+\kappa_0}\e^{-\sqrt{s/D}(x+x_0)}\right ).
\end{equation} 
It immediately follows that
\begin{equation}
\p(|x|,s|0)=\frac{1}{\sqrt{sD}+\kappa_0}\e^{-\sqrt{s/D}|x|},
\end{equation}
and, hence, equation (\ref{renewal3}) becomes
\begin{align}
  \label{renewal4}
 \widetilde{\rho}(x,s) = \p(x,s)+ \frac{ \kappa_0\e^{-\sqrt{s/D}|x|}}{ 2\sqrt{sD}}\Gamma(s),\ x\in {\mathbb G}.
 \end{align}
Note that in the limit $\kappa_0 \rightarrow 0$, we have $\widetilde{\rho}(x,s) \rightarrow \p(x,s)$. The fact that the particle may be found on either side of the barrier, even though it is now impenetrable, is simply an artifact of the initial distribution $g(x_0)$.

It follows from equation (\ref{renewal4}) that the density $\widetilde{\rho}(x,s)$ satisfies the Laplace transform of the semipermeable membrane BVP (\ref{class}) under the initial condition $\rho(x,0)=g(x)$ and $\kappa_0\rightarrow \kappa_0/2$. First, taking the second derivative of equations (\ref{renewal4}) for $x\neq 0^{\pm}$ and using equation (\ref{RobinLT}a) shows that
\begin{equation}
D\frac{\partial^2 \widetilde{\rho}(x,s)}{\partial x^2}-s \widetilde{\rho}(x,s)=-g(x), \quad x\in {\mathbb G}.
\end{equation}
Second, equation (\ref{renewal4}) implies that
\begin{subequations}
 \label{BCren}
 \begin{align}
  \widetilde{\rho}(x,s) + \widetilde{\rho}(-x,s)&= \p(x,s)+\p(-x,s)+ \frac{ \kappa_0\e^{-\sqrt{s/D}|x|}}{ \sqrt{sD}}\Gamma(s),\\
   \widetilde{\rho}(x,s) - \widetilde{\rho}(-x,s)&= \p(x,s)-\p(-x,s)
 \end{align}
 \end{subequations}
 for $x>0$. Differentiating equation (\ref{BCren}a) with respect to $x$ and taking $x=0^+$ we have
 \begin{align}
 \partial_x\widetilde{\rho}(0^+,s) - \partial_x\widetilde{\rho}(0^-,s) = \partial_x\p(0^+,s)- \partial_x\p(0^-,s)-\frac{\kappa_0}{D} \Gamma(s) .
 \end{align}
 The Robin boundary condition (\ref{RobinLT}b) implies that
 \begin{align*}
  \partial_x\p(0^+,s)- \partial_x\p(0^-,s)=\frac{\kappa_0}{D}[\p(0^+,s)+\p(0^-,s)]=\frac{\kappa_0}{D}\Gamma(s).
  \end{align*}
  Hence,
  \begin{equation}
  \label{L1}
 D \partial_x\widetilde{\rho}(0^+,s)=D\partial_x\widetilde{\rho}(0^-,s).
 \end{equation}
  Similarly, differentiating equation (\ref{BCren}b) with respect to $x$ and taking $x=0^+$ gives
 \begin{align}
 D\partial_x\widetilde{\rho}(0^+,s) +D \partial_x\widetilde{\rho}(0^-,s) &= D\partial_x\p(0^+,s)+ D\partial_x\p(0^-,s)\nonumber\\
 &=\kappa_0 [p(0^+,s)-p(0^-,s)]=\kappa_0 [\widetilde{\rho}(0^+,s)-\widetilde{\rho}(0^-,s)].
 \label{L2}
 \end{align}
Finally, combining equations (\ref{L1}) and (\ref{L2}) yields the permeable boundary condition
 \begin{equation}
 D \partial_x\widetilde{\rho}(0^{\pm},s)=\frac{\kappa_0}{2} [\widetilde{\rho}(0^+,s)-\widetilde{\rho}(0^-,s)].
 \label{ser}
 \end{equation}
  This establishes that the snapping out BM $X_t$ is the single-particle realization of the stochastic process whose probability density evolves according to the diffusion equation with a semipermeable membrane at $x=0$. It also follows that if $g(x_0)$ is an even function of $x_0$ then $\widetilde{\rho}(x,s)$ is an even function of $x$ so that the flux through the membrane is zero. In other words, it effectively acts as a totally reflecting barrier even though $\kappa_0>0$. It can also be checked that the solution of equation (\ref{renewal4}) reduces to
\begin{equation}
\widetilde{\rho}(x,s)=\frac{1}{4\sqrt{sD}}\left (\e^{-\sqrt{s/D}|x-x_0|}+\e^{-\sqrt{s/D}(x+x_0)}\right ), x>0.
\end{equation} 

 There are a number of reasons why it is advantageous to formulate diffusion through a semi-permeable barrier in terms of snapping out BM. First, it provides a method for simulating Brownian motion in the presence of such a barrier \cite{Lejay16}. Second, rather than solving a Fokker-Planck of the form (\ref{class}), we can express the (weak) solution for $\rho$ in terms of the solution $p$ of partially reflected BM. However, the major advantage within the context of the current paper is that it provides a powerful framework for developing more general probabilistic models of diffusion therough sempermeable membranes, as we illustrate in sections 3 and 4.

 \subsection{Thin-layer approximation}

It is instructive to relate the above probabilistic model of single-particle diffusion through a semi-permeable barrier to a recent study based on a Fokker-Planck description \cite{Kay22}. The latter was derived by taking a continuum limit of a continuous-time random walk model with a defect. Here we briefly show how the Fokker-Planck description is equivalent to using a thin-layer approximation of a semi-permeable barrier. In Ref. \cite{Lejay16} it is proven that the solution of the thin-layer BVP converges in distribution to the solution of the snapping out BM.

\begin{figure}[b!]
  \centering
  \includegraphics[width=8cm]{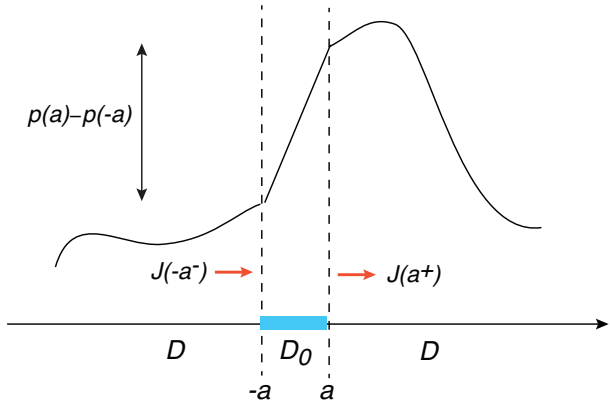}
  \caption{Thin layer problem. In the small $a$ limit, we have $J(-a^-)\approx J(a^+)\approx (\kappa_0/2) [p(a^+)-p(a^-)]$.}
  \label{fig1}
\end{figure}

In order to derive the thin-layer approximation, we first consider BM in ${\mathbb G}$ with a jump discontinuity in the diffusivity at $x=0$. That is, $D(x)=[D_+-D_-]H(x )+D_-$.
Introduce the stochastic process
\[X_t=F(X_t)\equiv\sqrt{2D_+}H(W_t)W_t+\sqrt{2D_-}H(-W_t)W_t.\]
Applying Ito's formula (\ref{Ito})
with
\[f'(x)=\sqrt{2D_+}H(x )+\sqrt{2D_-}H(-x),\ f''(x)= \sqrt{2D_+}\delta(x)-\sqrt{2D_-}\delta(-x),\]
yields the SDE
\[dX_t=[\sqrt{2D_+}H(W_t)+\sqrt{2D_-}H(-W_t)]dW_t+\frac{1}{2}[\sqrt{2D_+}\delta(W_t )-\sqrt{2D_-}\delta(-W_t)]dt.\]
Using $\mbox{sgn}(X_t)=\mbox{sgn}(W_t)$ and $\delta(\pm W_t)=\sqrt{2D_{\pm} }\delta(X_t)$ gives the skew BM \cite{Lejay06,App11,Lejay16}
\begin{align}
dX_t&=[\sqrt{2D_+}H(X_t)+\sqrt{2D_-}H(-X_t)]dW_t+[D_+-D_-]\delta(X_t)dt\nonumber \\
&=\sqrt{2D(X_t)}dW_t+\frac{D_+-D_-}{D_++D_-}dL_t^0(X),
\end{align}
where $L_t^0$ is the local time
\begin{equation}
L_t^0(X)=\frac{D_++D_-}{2}\int_0^t\delta(X_s)ds.
\end{equation}
The corresponding Ito FPE is then
\begin{align}
\frac{\partial p}{\partial t}
&= \frac{\partial}{\partial x} \frac{\partial D(x)p(x,t)}{\partial x}-\frac{\partial [D_+-D_-]\delta(x) p(x,t)}{\partial x}=\frac{\partial}{\partial x}\left [D(x) \frac{\partial p(x,t)}{\partial x}\right ].
\end{align}

Now consider the thin layer problem shown in Fig. \ref{fig1}. Outside the layer $[-a,a]$ the diffusivity is $D$, whereas within the layer it is $D_0$. Following from the previous calculation, we have the Ito SDE
\begin{equation}
dX_t=\sqrt{D(X_t)}dW_t+\frac{D-D_0}{D+D_0}dL_t^{a}(X)+\frac{D_0-D}{D+D_0}dL_t^{-a}(X),
\end{equation}
where $D(x)= D$ for $|x|>a$ and $D(x)=D_0$ for $|x|<a$, and the corresponding FPE
\begin{align}
\label{FPE}
\frac{\partial p}{\partial t}
&=\frac{\partial}{\partial x}\left [D(x) \frac{\partial p(x,t)}{\partial x}\right ].
\end{align}
Integrating the FPE across $x=-a$ and $x=+a$, respectively, yields the flux continuity conditions
\begin{equation}
D\partial_xp(-a^-,t)=D_0\partial_xp(-a^+,t),\quad D_0\partial_xp(a^-,t)=D\partial_xp(a^+,t).
\end{equation}
Suppose that $D_0=\kappa_0 a$ and consider the limit $a\rightarrow 0$. In the small-$a$ regime, we have
\begin{equation}
p(a,t)-p(-a,t)\approx 2a \partial_x p(-a^+,t)\approx 2a\partial_x p(a^-,t).
\end{equation}
Combining the various results gives, to leading order,
\[D\partial_xp(-a,t)\approx D\partial_xp(a,t)\approx \frac{D_0}{2a} [p(a,t)-p(-a,t)].\]
Finally, taking the limit $a\rightarrow 0^+$ recovers the permeable barrier boundary condition. Moreover, equation (\ref{FPE}) is equivalent to the FPE description derived in \cite{Kay22}.

\setcounter{equation}{0}
\section{Diffusion through a semipermeable membrane with stochastic resetting}

Let us return to the case of partially reflected BM in $[0,\infty)$, which is now supplemented by the resetting condition $X_t \rightarrow \xi\in [0,\infty)$ at a random sequence of times generated by a Poisson process with constant rate $r$. This particular problem has previously been studied in Refs.  \cite{Evans13,Bressloff22b}. The probability density $p_r(x,t|x_0)$ evolves according to the modified Robin BVP
\begin{subequations}
\begin{align}
\label{1Da}
 &\frac{\partial p_r}{\partial t}=D\frac{\partial^2p_r}{\partial x^2} -rp_r+r Q_r(x_0,t)\delta(x-\xi),\quad x>0,\\
&D\frac{\partial p_r}{\partial x}=\kappa_0 p_r,\quad x=0,\quad
 p_r(x,0|x_0) = \delta(x - x_0).
\label{1Db}
\end{align}
\end{subequations}
We have introduced the marginal distribution
\begin{equation}
\label{1DQ}
Q_r(x_0,t)=\int_0^{\infty} p_r(x,t|x_0)dx,
\end{equation}
which is the survival probability that the particle hasn't been absorbed at $x=0$ in the time interval $[0,t]$, having started at $x_0$. The $r$ subscript indicates the solution is in the presence of resetting. Note that in the limit $\kappa_0 \rightarrow 0$, the boundary at $x=0$ becomes totally reflecting so that $Q_r=1$ and we recover the standard forward equation for 1D diffusion with resetting \cite{Evans11a,Evans11b}. On the other hand, if $\kappa_0\rightarrow \infty$ then the boundary is totally absorbing. 

Laplace transforming equations (\ref{1Da}) and (\ref{1Db}) gives
\begin{subequations}
\begin{align}
\label{1DLTa}
  &D\frac{\partial^2\p_r(x,s|x_0)}{\partial x^2} -(r+s)p_r(x,s|x_0)=-[\delta(x-x_0)+r \Q_r(x_0,s)\delta(x-\xi),\, x>0,\\
 &D\frac{\partial \p_r(x,s|x_0)}{\partial x}=\kappa_0\p_r(x,s|x_0),\quad x=0.
\label{1DLTb}
\end{align}
\end{subequations}
Using the fact that $\p(x,s|x_0)$ is the Green's function for partially reflecting BM without resetting, see equation (\ref{EBM}), it follows that
  \begin{eqnarray}
\label{p1D}
   \p_r(x, s|x_0) =  \p(x,r+s|x_0)+ r\Q_r(x_0,s)\p(x,r+s|\xi), \quad 0<x<\infty.
\end{eqnarray}
 Finally, Laplace transforming equation (\ref{1DQ}) and using (\ref{p1D}) shows that
 \begin{align}
  \Q_r(x_0,s)&=\int_0^{\infty}\p_r(x,s|x_0)dx\nonumber \\
  &=\int_0^{\infty} \p(x,r+s|x_0)dx+  r\Q_r(x_0,s)\int_0^{\infty} \p(x,r+s|\xi)dx\nonumber \\
&=\Q(x_0,r+s)+r\Q_r(x_0,s)\Q(\xi,r+s),
\label{QQr}
 \end{align}
 where $\Q$ is the Laplace transform of the survival probability without resetting:
 \begin{equation}
 \label{QQ0}
 \Q(x_0,s)=\frac{1-\e^{-\sqrt{s/D}x_0}}{s}+\frac{\e^{-\sqrt{s/D}x_0}}{s+\kappa_0\sqrt{s/D}}.
 \end{equation}
  Rearranging equation (\ref{QQr}) thus determines the survival probability with resetting in terms of the corresponding probability without resetting:
 \begin{equation}
 \label{Qr}
 \Q_r(x_0,s)=\frac{\Q(x_0,r+s)}{1-r\Q(\xi,r+s)}.
 \end{equation}
 For $\kappa_0>0$ the steady-state survival probability vanishes with or without resetting, since 1D diffusion is recurrent so that absorption eventually occurs. Indeed,
 \begin{align}
 Q_r^*(x_0)&=\lim_{s\rightarrow 0}s\Q_r(x_0,s) =\lim_{s\rightarrow 0}\frac{s\Q(x_0,r)}{1-r\Q(\xi,r)}=0.
 \end{align}
 (Note that $\Q(\xi,r)\neq 1/r$ when $\kappa_0>0$.) On the other hand, if $\kappa_0=0$ (totally reflecting boundary at $x=0$), then $\Q_r(x_0,s) =1/s $ for all $x_0<\infty$ and thus $Q_r^*(x_0)=1$. In this special case, there exists a nonequilibrium stationary state (NESS) given by
 \begin{align}
  p_r^*(x)&=\lim_{s\rightarrow 0}s\p_r(x,s|x_0)= \lim_{s\rightarrow 0}s[\p(x,r+s|x_0)+r\Q_r(x_0,s)\p(x,r+s|\xi)]\nonumber \\
  &=r\p(x,r|\xi)=\frac{r}{2\sqrt{rD}}\left [\e^{-\sqrt{r/D} |x-\xi|}+\e^{-\sqrt{r/D} |x+\xi|}\right ],\quad x>0,
  \label{ref}
 \end{align}
 which recovers the well-known result of Refs. \cite{Evans11a,Evans11b}.
 
 \begin{figure}[t!]
  \centering
  \includegraphics[width=8cm]{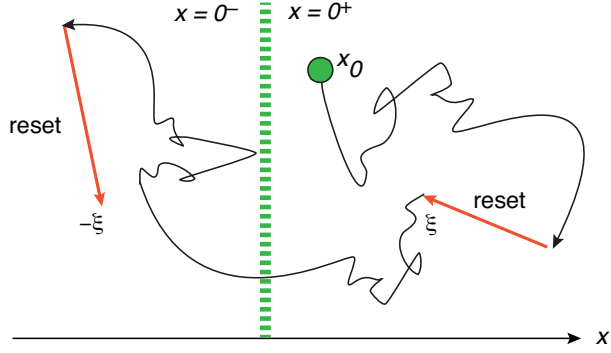}
  \caption{Single-particle diffusion through a semipermeable membrane with stochastic resetting to $\pm \xi$. (The dynamics is extended into two dimensions for ease of visualization.) The snapped out BM starts on the right-hand side of the membrane, say, and undergoes one reset to $+\xi$ before passing through the membrane to the left-hand side. Whilst in this domain the particle resets to $-\xi$ and so on. Resetting events that cross the membrane are forbidden.}
  \label{fig2}
\end{figure}

We now observe that partially reflecting BM with resetting is also a strong Markov process, since there is no memory of previous histories following resetting to $\xi$. This means that a modified version of the renewal equation (\ref{renewal3}) for snapping out BM holds when resetting is included. For simplicity, suppose that 
we sew together positively and negatively reflecting BMs such that the former resets to $\xi$ and the latter to $-\xi$ with $\xi\geq 0^+$, see Fig. \ref{fig2}. This symmetric resetting protocol means that $p_r(x,s)=p_r(-x,s)$. It follows that the renewal equation (\ref{renewal3}) becomes\footnote{We could consider a more general resetting protocol in which $X_t\rightarrow \xi_+>0$ when $X_t\geq 0^+$ and $X_t\rightarrow \xi_-<0$ when $X_t\leq 0^-$ with $|\xi_-|\neq \xi_+$ by an appropriate modification of the renewal equation.}
\begin{align}
  \label{reset1}
 \widetilde{\rho}_r(x,s) = \p_r(x,s)+ \frac{ \kappa_0\Gamma_r(s)/2}{1-\kappa_0\p_r(0,s|0)} \p_r(|x|,s|0) ,\ x\in {\mathbb G},\ \kappa_0>0
 \end{align}
 with $\Gamma_r(s)=\p_r(0^+,s)+\p_r(0^-,s)$. Note that our resetting protocol is space-dependent due to the fact that we exclude resetting events that involve a particle crossing the semipermeable membrane to the other side. Hence, spatial position $X_t \geq 0^+$ ($X_t\leq 0^-$) can only reset to $x=\xi$ ($x=-\xi$). (Most models of stochastic resetting take resetting to be independent of the current location $X_t$ \cite{Evans20}. Examples of space-dependent resetting protocols can be found in Refs. \cite{Evans11b,Roldan17,Pinsky20}.) This means that we have to work with the modified renewal equation (\ref{reset1}) that keeps track of which side of the membrane a particle is located, rather than using a renewal equation that directly relates $\rho_r(x,t)$ to $\rho(x,t)$. In other words, we cannot simply introduce the resetting protocol into the Fokker-Planck equation for $\rho_r(x,t)$.
 
 \subsection{Nonequilibrium stationary state}
 
 One of the common characteristic features of non-absorbing diffusion processes with stochastic resetting is that there exists a nonequilibrium stationary state (NESS), which is maintained by non-zero probability fluxes \cite{Evans20}. In the case of snapping out BM with resetting, the points $x=\pm \xi$ act as probability sources, whereas all positions $x\neq \pm \xi$ are potential probability sinks. Although each partially reflected BM is killed by absorption at the semipermeable barrier, the stochastic process is immediately restarted so that snapping out BM is not killed. We will derive the NESS using the renewal equation (\ref{reset1}). Multiplying both sides by $s$ and taking the limit $s\rightarrow 0$ gives
 \begin{align}
 \rho_r^*(x)&=\lim_{t\rightarrow \infty}\rho_r(x,t)=\lim_{s\rightarrow 0}s\widetilde{\rho}_r(x,s)\nonumber \\
 &=\frac{\kappa_0}{2} \lim_{s\rightarrow \infty}\frac{s\Gamma_r(s)}{1-\kappa_0\p_r(0,s|0)} \p_r(|x|,s|0) .
 \end{align}
We have used the fact that partially reflected BM with resetting does not have a nontrivial NESS, that is, $\lim_{t\rightarrow \infty}p_r(x,t)=0$. The existence of the NESS for snapping out BM can be established by showing that $1-\kappa_0 p_r(0,s|0)=O(s)$. Setting $x=x_0=0$ in equation (\ref{p1D}) and using equation (\ref{EBM}) yields
  \begin{align}
\label{p1D2}
   \p_r(0, s|0) &=  \p(0,r+s|0)+ r\Q_r(0,s)\p(0,r+s|\xi)\nonumber \\
   &=\frac{1}{\sqrt{(r+s)D}+\kappa_0}\left [1+  r\e^{-\sqrt{(r+s)/D}\xi}\Q_r(0,s)\right ].
   \end{align}
Equations (\ref{QQ0}) and (\ref{Qr}) give
\begin{align}
\Q_r(0,s)&=\frac{1 \slash  [r+s+\kappa_0\sqrt{(r+s)/D}] }{1-\frac{\displaystyle r}{\displaystyle r+s}\left (1-\e^{-\sqrt{(r+s)/D}\xi} \right )-\frac{\displaystyle r}{\displaystyle r+s+\kappa_0 \sqrt{[r+s]/D}}\e^{-\sqrt{(r+s)/D}\xi}}\nonumber \\
&=\frac{r+s}{s[r+s+\kappa_0\sqrt{(r+s)/D}] +\kappa_0 r\sqrt{(r+s)/D}\e^{-\sqrt{(r+s)/D}\xi} }.
\end{align}
Substituting into equation (\ref{p1D2}) and rearranging, we find that
\begin{align}
 \p_r(0, s|0) &=\frac{1}{\sqrt{(r+s)D}+\kappa_0}\nonumber \\
 &\quad \times \left [1+\frac{1}{\kappa_0} \frac{\sqrt{(r+s)D}}{1+\frac{\displaystyle s}{\displaystyle r\kappa_0}[r+s+\kappa_0\sqrt{(r+s)/D}]\sqrt{\frac{\displaystyle D}{\displaystyle r+s}} \e^{\sqrt{(r+s)/D}\xi} } \right ]\nonumber \\
 &=\frac{1}{\sqrt{(r+s)D}+\kappa_0}\nonumber \\
 &\quad \left [1+\frac{\sqrt{(r+s)D}}{\kappa_0} \left (1-\frac{\displaystyle s [r+\kappa_0\sqrt{r/D}]}{\displaystyle r\kappa_0}\sqrt{\frac{\displaystyle D}{\displaystyle r}} \e^{\sqrt{r/D}\xi}+O(s^2) \right ) \right ]\nonumber \\
& = \frac{1}{\kappa_0}\left [1-\frac{s}{\kappa_0}\sqrt{\frac{\displaystyle D}{\displaystyle r} }  \e^{\sqrt{r/D}\xi}\right ]+O(s^2)
 \end{align}
It immediately follows that $1-\kappa_0 \p_r(0, s|0)=O(s)$ and thus
\begin{align}
 \rho_r^*(x)&=\frac{\kappa_0^2}{2}\sqrt{\frac{\displaystyle r}{\displaystyle D} }  \e^{-\sqrt{r/D}\xi} \Gamma_r(0) \p_r(|x|,0|0) .
 \end{align}
 
 \begin{figure}[t!]
  \centering
  \includegraphics[width=10cm]{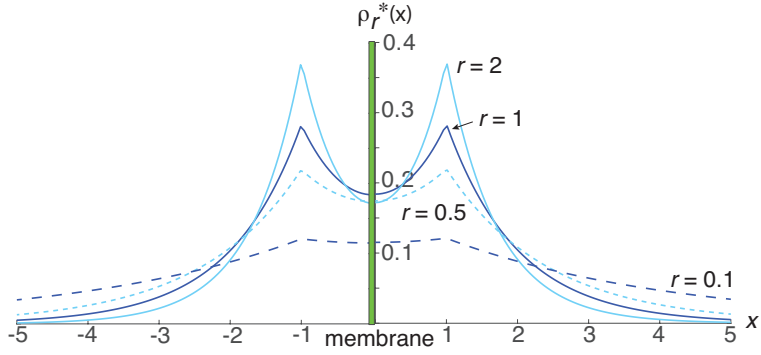}
  \caption{NESS for the snapping out BM with resetting. The density $\rho_r^*(x)$ is plotted as a function of $x$ for various resetting rates $r$ and $\xi=1$. We also set $D=1$.}
  \label{fig3}
\end{figure}
 
 The factor $\Gamma_r(0)$ is
 \begin{align}
 \Gamma_r(0)&=\p_r(0^+,0)+\p_r(0^-,0)=\int_{-\infty}^{\infty}g(x_0)[ \p(0,r||x_0|)+ r\Q_r(|x_0|,0)\p(0,r|\xi)]dx_0\nonumber\\
 &=\frac{1}{\sqrt{rD}+\kappa_0}\int_{-\infty}^{\infty}g(x_0)\left [\e^{-\sqrt{r/D}|x_0|}+\frac{r\Q(|x_0|,r)}{1-r\Q(\xi,r)} \e^{-\sqrt{r/D}\xi}\right ]\nonumber \\
 &=\frac{1}{\kappa_0}\int_{-\infty}^{\infty}g(x_0)dx_0=\frac{1}{\kappa_0}.
 \end{align}
Hence, the NESS takes the form
\begin{equation}
 \rho_r^*(x)=\frac{\kappa_0}{2}\sqrt{\frac{\displaystyle r}{\displaystyle D} }  \e^{-\sqrt{r/D}\xi}\p_r(|x|,0|0).
 \end{equation}
 As expected, $\rho_r^*(x)$ is independent of the initial distribution $g(x_0)$ and is an even function of $x\in {\mathbb G}$.
 Finally, combining equations (\ref{p1D}), (\ref{QQ0}) and (\ref{Qr}) shows that
 \begin{align}
 \p_r(x,0|0)&=p(x,r|0)+\frac{\sqrt{r D}}{\kappa_0}\e^{\sqrt{r/D}\xi} \p(x,r|\xi)\nonumber \\
 &=\frac{1}{\sqrt{rD}+\kappa_0}\e^{-\sqrt{r/D}x}\nonumber \\
 &\quad +\frac{\e^{\sqrt{r/D}\xi} }{2\kappa_0}\left (\e^{-\sqrt{r/D}|x-\xi|}+\frac{\sqrt{rD}-\kappa_0}{\sqrt{rD}+\kappa_0}\e^{-\sqrt{r/D}(x+\xi)}\right )\nonumber \\
 &=\frac{1}{2\kappa_0}\left  ( \e^{-\sqrt{r/D}x}+\e^{\sqrt{r/D}\xi}\e^{-\sqrt{r/D}|x-\xi|}\right )
 \end{align}
 and, hence (see Fig. \ref{fig3})
 \begin{equation}
 \rho_r^*(x)=\frac{r}{2}\frac{1}{2\sqrt{rD} }\left  ( \e^{-\sqrt{r/D}(x+\xi)}+ \e^{-\sqrt{r/D}|x-\xi|}\right )=\frac{p_r^*(|x|)}{2},
 \end{equation}
 where we have used equation (\ref{ref}). Note that the NESS is independent of $\kappa_0$ for $\kappa_0>0$ and has the following interpretation. In the long time limit, the particle spends an equal amount of time on either side of the barrier where it undergoes repeated rounds of partially reflecting BM with resetting. Thus each side forms the NESS $p_r^*(|x|)$ but is weighted by a factor of $1/2$. The limit $\kappa_0 \rightarrow 0$ is singular, since the relative weight of the density on either side of the barrier will depend on the initial density $g(x_0)$.

 \subsection{Relaxation time}
 
 Although $\rho_r^*(x)$ is independent of the permeability $\kappa_0$, the time-dependent relaxation to the NESS will be $\kappa_0$-dependent. In the case of homogeneous diffusion in $\R^d$, one can use large deviation theory to show that the approach to the stationary state exhibits a dynamical phase transition, which can be interpreted as a traveling front separating spatial regions for which the probability density has relaxed to the NESS from those where transients persist \cite{Majumdar15}. Recently, we introduced an alternative method for characterizing the relaxation process, which is based on the notion of an accumulation time \cite{Bressloff21a}. We proceeded by decomposing the probability density into decreasing and accumulating components, and showed how the latter evolved in an analogous fashion to the formation of a concentration gradient in diffusion-based morphogenesis. The accumulation time for the latter is the analog of the mean first passage time of a search process, in which the survival probability density is replaced by an accumulation fraction density \cite{Berez10,Berez11,Gordon11}.

 Following Ref. \cite{Bressloff21a}, consider the function
\begin{equation}
\label{Z}
Z_r(x,t)=1-\frac{\rho_r(x,t)}{\rho_r^*(x)},
\end{equation}
and define 
\begin{equation}
T_r(x)=\int_0^{\infty} Z_r(x,t)dt=\lim_{s\rightarrow 0} \widetilde{Z}_r(x,s) .
\end{equation}
 Laplace transforming equation (\ref{Z}) gives
\[\widetilde{Z}_r(x,s)=\frac{1}{s}\left [1-\frac{s\widetilde{\rho}_r(x,s)}{\rho_r^*(x)}\right ]
\]
and, hence,
\begin{align}
T_r(x)&= \lim_{s\rightarrow 0}\frac{1}{s}\left [1-\frac{s\widetilde{\rho}_r(x,s)}{\rho_r^*(x)}\right ] =-\frac{1}{\rho_r^*(x)}
\left .\frac{d}{ds}[s\widetilde{\rho}_r(x,s)]\right |_{s=0}.
\label{acc}
\end{align}
We have used the identity $\rho_r^*(x)=\lim_{s\rightarrow 0}s\widetilde{\rho}_r(x,s)$. In cases where $Z_r(x,t)$ is a positive function of $x$ for all $t>0$ (no overshooting), we can interpret $T_r(x)$ as the mean accumulation time to the stationary state. However, positivity of $Z_r(x,t)$ does not necessarily hold in the case of stochastic processes with resetting. Nevertheless, as shown in Ref. \cite{Bressloff21}, one can decompose $T_r(x)$ into negative and positive parts and interpret the latter as an accumulation time. Since the first term in equation (\ref{reset1}) does not contribute to the NESS and generates a negative contribution to $T(x)$, we define the accumulation time as
\begin{equation}
T^*(x)=-\frac{1}{\rho_r^*(x)}
\left .\frac{d}{ds}[s(\widetilde{\rho}_r(x,s)-\p_r(x,s))]\right |_{s=0}.
\end{equation}

 \begin{figure}[t!]
  \centering
  \includegraphics[width=10cm]{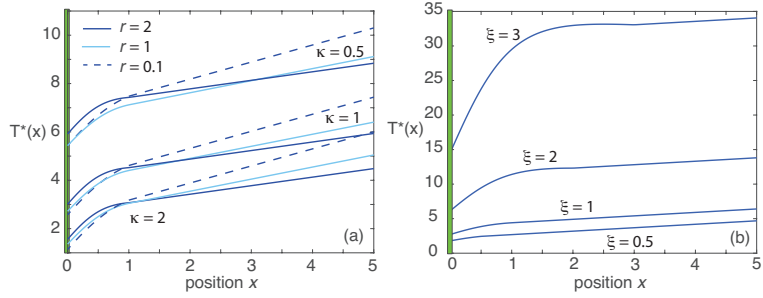}
  \caption{Accumulation time for the snapping out BM with resetting. (a) $T_r^*(x)$ is plotted as a function of $x$ for various resetting rates $r$ and absorption rates $\kappa_0$ with $\xi=1$. We also set $D=1$ and $g(x_0)=\delta(x-x_0)$ with $x_0=1$. (b) Corresponding plots for various resetting positions $\xi$ with $\kappa_0=1$and $r=1$.}
  \label{fig4}
\end{figure}

\begin{figure}[b!]
  \centering
  \includegraphics[width=8cm]{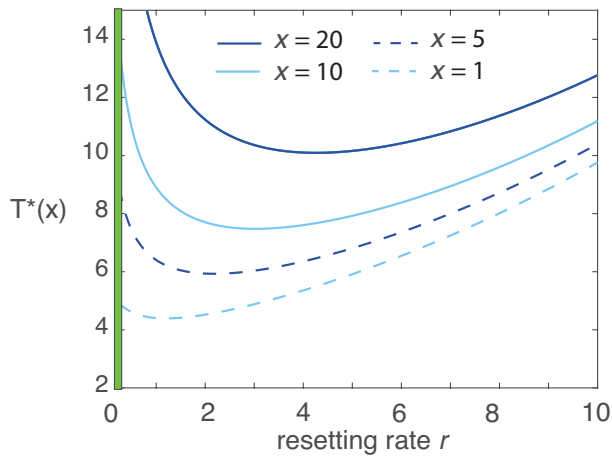}
  \caption{Accumulation time for the snapping out BM with resetting. $T_r^*(x)$ is plotted as a function of $r$ for various spatial locations $x$. We also set $D=1$, $\xi=1$, $\kappa_0=1$ and $g(x_0)=\delta(x-x_0)$ with $x_0=1$.}
  \label{fig5}
\end{figure}

In Fig. \ref{fig4} we plot $T^*(x)$ as a function of $x$, $x>0$, for various choices of model parameters and the initial condition $g(x_0)=\delta(x-x_0)$. A number of observations can be made. First, $T^*(x)$ for fixed $x$ is a decreasing function of $\kappa_0$ and an increasing function of $\xi$, which reflects the fact that each round of partially reflected BM takes longer on average. Second, there is a cross-over phenomenon whereby $T^*(x) $ is a non-monotonic function of the resetting rate $r$ for fixed $x$. This is further illustrated in Fig. \ref{fig5}, which indicates that. $T^*(x)$ for fixed $x$ is a unimodal function of $r$ with a minimum at an $x$-dependent rate $r^*(x)$. Third, $T^*(x)$ asymptotically approaches a linear function of $x$, which is consistent with previous findings in other systems \cite{Majumdar15,Bressloff21a}. Finally, note that if we had considered $T(x)$ rather than $T^*(x)$ then $T(x)$ would be negative for locations close to the membrane. 

  \setcounter{equation}{0}
 \section{Encounter-based version of snapping out BM}
 
Another possible extension of snapping out BM is to modify the rule for killing each round of partially reflected BM. This is equivalent to changing the absorption process on either side of the semipermeable barrier. We proceed by using the so-called encounter-based model of absorption \cite{Grebenkov20,Grebenkov22,Bressloff22,Bressloff22a}, which replaces the exponential distribution for the stopping local time threshold $\widehat{\ell}$, see equation (\ref{exp}), by a non-exponential distribution. The basic idea is to introduce 
the joint probability density or generalized propagator for the pair $(X_t,L_t)$, where $X_t\in [0,\infty)$ is partially reflected BM and $L_t$ is the local time at $x=0$:
\[P(x,\ell,t|x_0)dx\, d\ell :=\P[x\leq X_t <x,x+dx, \ell \leq  L_t<\ell+d\ell|X_0=x_0,\ell_0=0].\]
Since the local time only changes at the membrane boundary $x=0$, the evolution equation within the bulk of the domain is simply 
 \begin{align} 
   \label{JPCK1}
   \frac{\partial P}{\partial t}  = D\frac{\partial^2 P}{\partial x^2},\ x>0, \ \ell \geq 0, \ t>0.
     \end{align}
The nontrivial step is determining the boundary condition at $x=0$. Here we give a heuristic derivation that considers a thin layer in a neighborhood of the boundary given by the interval $[0,h]$ with
\begin{align}
L_t^h=\frac{D}{h} \int_0^t\left [\int_{0}^h\delta(X_{t'}-x)dx\right ]dt' .
\end{align}
By definition, $hL_t^h$ is the residence or occupation time of the process $X_t$ in the boundary layer $[0,h]$ up to time $t$.
Although the width $h$ and the residence time in the boundary layer vanish in the limit $h\rightarrow 0$, the rescaling by $1/h$ ensures the nontrivial limit
$L_t=\lim_{h\rightarrow 0} L_t^h$.
Moreover, from conservation of probability, the flux into the boundary layer over the residence time $h\delta \ell$ generates a corresponding shift in the probability $P$ within the boundary layer from $\ell\rightarrow \ell +\delta \ell$. That is, for $\ell>0$,
\begin{eqnarray*}
-J(h,\ell,t|x_0)h\delta \ell =[P(0,\ell+\delta\ell,t|x_0)-P(0,\ell,t|x_0)]h ,
\end{eqnarray*}
where $J(x,\ell,t|x_0)=-D\partial_x P(x,\ell,t|x_0)$.
Dividing through by $h\delta \ell$ and taking the limits $h\rightarrow 0$ and $\delta \ell \rightarrow 0$ yields 
 $-J(0,\ell,t|x_0)=\partial_{\ell}P(0,\ell,t|x_0)$, $\ell >0$.
Moreover, when $\ell=0$ the probability flux $J(0,0,t|x_0)\delta \ell$ is identical to that of a Brownian particle with a totally absorbing boundary at $x=0$, which we denote by $J_{\infty}(0,t|x_0)$. Combining all of these results yields the boundary condition 
\begin{equation}
\label{BCL}
-J(0,\ell,t|x_0)=-J_{\infty}(0,t|x_0)\delta(\ell)+\frac{\partial P(0,\ell,t|x_0)}{\partial \ell}.
\end{equation}
It can also be shown that $P(0,0,t|x_0)=-J_{\infty}(L,t|x_0)$.
 For a more detailed derivation of the boundary condition (\ref{BCL}) see Refs. \cite{Grebenkov20,Bressloff22}.
Finally, Laplace transforming  equations (\ref{JPCK1}) and (\ref{BCL}) with respect to $\ell$ 
by setting
\begin{equation}
\PP(x,z,t|x_0)=\int_0^{\infty}\e^{-z\ell} P(x,\ell,t|x_0)d\ell
\end{equation} 
we find that the $\PP(x,z,t|x_0)$ is the solution to the Robin BVP (\ref{Robin}) with $\kappa_0=D z$ and $z$ the Laplace variable.

The above is consistent with the observation that partially reflected BM is obtained by supplementing reflected BM
with a stopping condition that halts the stochastic process when the local time $L_t(X)$ exceeds a random exponentially distributed threshold $\widehat{\ell}$. This can be established as follows. Given that $L_t$ is a nondecreasing process, the condition $t < {\mathcal T}$ is equivalent to the condition $L_t <\widehat{\ell}$. This implies that 
\begin{align*}
p(x,t|x_0)dx&=\P[x\leq X_t <x+dx, \ L_t < \widehat{\ell}|X_0=x_0]\\
&=\int_0^{\infty} d\ell \ \psi(\ell) \P[x\leq X_t <x+dx, \ L_t < \ell |X_0=x_0]\\
&=\int_0^{\infty} d\ell \psi(\ell)\int_0^{\ell} d\ell' [P(x,\ell',t|x_0)dx],
\end{align*}
where $\psi(\ell)=-\Psi'(\ell)=(\kappa_0/D) \e^{-\kappa_0 \ell/D}$.
Using the identity
\[\int_0^{\infty}d\ell \ u(\ell)\int_0^{\ell} d\ell' \ v(\ell')=\int_0^{\infty}d\ell' \ v(\ell')\int_{\ell'}^{\infty} d\ell \ u(\ell)\]
for arbitrary integrable functions $u,v$, it follows that
\begin{align}
\label{pP}
p(x,t|x_0)&=\int_0^{\infty}P(x,\ell',t|x_0)\left [\int_{\ell'}^{\infty} \psi(\ell)d\ell\right ]d\ell'  
=\int_0^{\infty}\Psi(\ell) P(x,\ell,t|x_0)d\ell .
\end{align}
Hence, the probability density of partially reflected BM is equivalent to the Laplace transform of the local time propagator with $z=\kappa_0/D$ acting as the Laplace variable. Assuming that the Laplace transform can be inverted, we can then incorporate a non-exponential probability distribution $\Psi(\ell)$ such that the corresponding marginal density is
  \begin{equation}
  \label{oo}
  p_{\Psi}(x,t|x_0)=\int_0^{\infty} \Psi(\ell)P(x,\ell,t|x_0)d\ell = \int_0^{\infty} \Psi(\ell){\mathcal L}_{\ell}^{-1}\widetilde{P}(x,z,t|x_0)d\ell.
  \end{equation}
  One major difference from the exponential law $\Psi(\ell)=\e^{-\kappa_0/D}$ is that the stochastic process $X_t$ is no longer Markovian. One way to see this is to note that a non-exponential distribution can be generated by an $\ell$-dependent absorption rate, $\kappa=\kappa(\ell)$. That is,
  \begin{equation}
  \Psi(\ell)=\exp(-D^{-1}\int_0^{\ell}\kappa(\ell')d\ell').
  \label{kapl}
  \end{equation}
  Given that the probability of absorption now depends on how much time the particle spends in a neighborhood of the boundary, as specified by the local time, it follows that the stochastic process has memory.

We now define a generalized snapping out BM as follows. Again we assume that the particle starts at $X_0=x_0\geq 0$. It realizes positively reflected BM until its local time $L_t$ at $x=0^+$ is greater than an independent random variable $\widehat{\ell}$ with a nonexponential distribution $\Psi(\ell)=\P[\widehat{\ell} > \ell]$. It then randomly determines its sign with probability 1/2 and restarts as a new reflected BM in either $[0^+,\infty)$ or $(-\infty,0^-]$, and so on. Although each round of partially reflected Brownian motion is non-Markovian, all history is lost following absorption and restart so that we can construct a renewal equation. However, it is now more convenient to use a first rather than a last renewal equation.

Let $p_{\Psi}(x,t)$ denote the extended probability density on $x\in {\mathbb G}$ with
\begin{equation}
p_{\Psi}(x,t)=H(x)\int_{0}^{\infty}p_{\Psi}(x,t|x_0)g(x_0)dx_0+H(-x)\int_{-\infty}^{0}p_{\Psi}(-x,t|-x_0)g(x_0)dx_0,
\end{equation}
where $p_{\Psi}(x,t|x_0)$ for $x,x_0\geq 0$ is the generalized partially reflecting BM.
 Let $Q_{\Psi}(t)$ denote the corresponding survival probability 
 \begin{equation}
 Q_{\Psi}(t)=\int_{-\infty}^{\infty} p_{\Psi}(x,t)dx.
 \end{equation}
 It follows that the first passage time density for absorption is $f_{\Psi}(t)=-dQ_{\Psi}(t)/dt$. The first renewal equation then takes the form
 \begin{align}
  \label{Frenewal}
 \rho_{\Psi}(x,t)&=p_{\Psi}(x,t)+\frac{1}{2}\int_0^t [\rho_{\Psi}(x,t-\tau |0^+) +\rho_{\Psi}(x,t-\tau|0^- )]f_{\Psi}(\tau)d\tau ,\ x\in {\mathbb G}.
   \end{align}
 The first term on the right-hand side represents all sample trajectories that have never been absorbed by the barrier at $x=0^{\pm}$ up to time $t$. The corresponding integrand represents all trajectories that were first absorbed (stopped) at time $\tau$ and then switched to either positively or negatively reflected BM state with probability 1/2, after which an arbitrary number of switches can occur before reaching $x$ at time $t$. The probability that the first stopping event occurred in the interval $(\tau,\tau+d\tau)$ is $f_{\Psi}(\tau) d\tau$. Laplace transforming the renewal equation (\ref{Frenewal}) with respect to time $t$ by setting $\widetilde{\rho}_{\Psi}(x,s) =\int_0^{\infty}\e^{-st}\rho_{\Psi}(x,t)dt$ etc. gives
 \begin{align}
  \label{Frenewal2}
 \widetilde{\rho}_{\Psi}(x,s) = \p_{\Psi}(x,s)+\frac{1}{2}[\widetilde{\rho}_{\Psi}(x,s |0^+)+\widetilde{\rho}_{\Psi}(x,s |0^-) ]\widetilde{f}_{\Psi}(s),\ x\in {\mathbb G}.
 \end{align}
 Moreover,
$ \widetilde{f}_{\Psi}(s)=1-s\widetilde{Q}_{\Psi}(s)$. In order to determine the factor $\widetilde{\rho}_{\Psi}(x,s |0^+)+\widetilde{\rho}_{\Psi}(x,s |0^-)$ we set $g(x_0)=[\delta(x_0-0^+) +\delta(x-0^-)]/2$ in equation (\ref{Frenewal2}). This gives
   \begin{align*}
\widetilde{\rho}_{\Psi}(x,s |0^+)+\widetilde{\rho}_{\Psi}(x,s |0^-)=2\p_{\Psi}(|x|,s|0)+[\widetilde{\rho}_{\Psi}(x,s |0^+)+\widetilde{\rho}_{\Psi}(x,s |0^-) ]\widetilde{f}_{\Psi}(0,s),
 \end{align*}
 which can be arranged to obtain the result
    \begin{align*}
\widetilde{\rho}_{\Psi}(x,s |0^+)+\widetilde{\rho}_{\Psi}(x,s |0^-)= \frac{2\p_{\Psi}(|x|,s|0)}{s\widetilde{Q}_{\Psi}(0,s)}.
 \end{align*}
  Substituting back into equations (\ref{Frenewal2}) yields the explicit solution
 \begin{align}
  \label{Frenewal3}
 \widetilde{\rho}_{\Psi}(x,s) = \p_{\Psi}(x,s)+ \frac{1-s\Q_{\Psi}(s)}{s\widetilde{Q}_{\Psi}(0,s)}\p_{\Psi}(|x|,s|0) ,\ x\in {\mathbb G}.
 \end{align}
It can be checked that equations (\ref{renewal3}) and (\ref{Frenewal3}) agree when $\Psi(\ell)= \e^{-\kappa_0 \ell/D}$ so that $\p_{\Psi}(x,s|x_0) \rightarrow \p(x,s|x_0)$ and $\Q_{\Psi}(x_0,s) \rightarrow \Q(x_0,s)$ with $\p$ and $\Q$ given by equations (\ref{EBM}) and (\ref{QQ0}), respectively.
Indeed, since
\[\p(0,s|x_0)=\frac{1}{\sqrt{sD}+\kappa_0}\e^{-\sqrt{s/D}x_0},\]
we see that $1-s\Q(x_0,s)=\kappa_0 \p(0,s|x_0)$ and, hence, 
\[1-s\Q(s)=\kappa_0 \Gamma(s)/2,\quad s\Q(0,s)=1-\kappa_0 \p(0,s|0).\]

It remains to calculate $\p_{\Psi}$. From equation (\ref{EBM}) we have
\begin{align}
\label{EBM2}
\calP(x,z,s|x_0)&\equiv \int_0^{\infty}\e^{-st}\left [\int_0^{\infty} \e^{-z\ell}P(x,\ell,t|x_0)d\ell\right ]dt\nonumber \\
&=\frac{1}{2\sqrt{sD}}\left (\e^{-\sqrt{s/D}|x-x_0|}+\frac{\sqrt{sD}-Dz}{\sqrt{sD}+Dz}\e^{-\sqrt{s/D}(x+x_0)}\right ).
\end{align} 
Inverting the Laplace transform in $z$ gives
\begin{align}
\PP(x,\ell,s|x_0)&=\frac{1}{2\sqrt{sD}}\left (\e^{-\sqrt{s/D}|x-x_0|}-\e^{-\sqrt{s/D}(x+x_0)}  \right )\delta(\ell)\nonumber \\
&\quad +\frac{1}{D}\e^{-\sqrt{s/D}(x+x_0)} \e^{-\sqrt{s/D}\ell}  .
\label{Ptil}
\end{align}
Substituting into equation (\ref{oo}) after Laplace transforming the latter with respect to $t$, we obtain the result
\begin{align}
\p_{\Psi}(x,s|x_0)&=\frac{1}{2\sqrt{sD}}\left (\e^{-\sqrt{s/D}|x-x_0|}-\e^{-\sqrt{s/D}(x+x_0)}  \right )\nonumber \\
&\quad  +\frac{1}{D}\e^{-\sqrt{s/D}(x+x_0)}\widetilde{\Psi}(\sqrt{s/D}).
\end{align}
It immediately follows that
\begin{align}
\p_{\Psi}(x,s|0)=\p_{\Psi}(0,s|x)=\frac{1}{D}\e^{-\sqrt{s/D}x}\widetilde{\Psi}(\sqrt{s/D})
\end{align}
and
\begin{equation}
 \label{QQpsi}
 \Q_{\psi}(x_0,s)=\frac{1-\e^{-\sqrt{s/D}x_0}}{s}+\frac{\e^{-\sqrt{s/D}x_0}}{\sqrt{sD}}\widetilde{\Psi}(\sqrt{s/D}).
 \end{equation}
 Hence, equation (\ref{Frenewal3}) reduces to the form
 \begin{align}
  \label{Frenewal4}
 \widetilde{\rho}_{\Psi}(x,s) = \p_{\Psi}(x,s)+ \frac{\e^{-\sqrt{s/D}|x|}}{ 2\sqrt{sD}}\Gamma_{\Psi}(s),\ x\in {\mathbb G},
 \end{align}
 where
 \begin{align}
 \Gamma_{\Psi}(s)=\left [1-\sqrt{\frac{s}{D}}\widetilde{\Psi}(\sqrt{s/D}) \right ]\int_{-\infty}^{\infty}\e^{-\sqrt{s/D}|x_0|}f(x_0)dx_0.
\end{align}

Since the propagator satisfies the diffusion equation in the bulk of the domain, the density $\rho_{\Psi}(x,t)$ does too. The remaining issue concerns the boundary condition at the interface. Using similar arguments to section 2, equations (\ref{BCren})--(\ref{L2}), we find
\begin{subequations}
 \label{enc0}
 \begin{align}
  \widetilde{\rho}_{\Psi}(x,s) + \widetilde{\rho}_{\Psi}(-x,s)&= \p_{\Psi}(x,s)+\p_{\Psi}(-x,s)+ \frac{ \e^{-\sqrt{s/D}|x|}}{ D}\Gamma_{\Psi}(s),\\
   \widetilde{\rho}_{\Psi}(x,s) - \widetilde{\rho}_{\Psi}(-x,s)&= \p_{\Psi}(x,s)-\p_{\Psi}(-x,s),
 \end{align}
 \end{subequations}
 and
\begin{subequations}
\label{enc1}
 \begin{align}
&D \partial_x\widetilde{\rho}_{\Psi}(0^+,s) - D\partial_x\widetilde{\rho}_{\Psi}(0^-,s) = D\partial_x\p_{\Psi}(0^+,s)- D\partial_x\p_{\Psi}(0^-,s)- \Gamma_{\psi}(s) ,\\
& D\partial_x\widetilde{\rho}_{\Psi}(0^+,s) +D \partial_x\widetilde{\rho}_{\Psi}(0^-,s) = D\partial_x\p_{\Psi}(0^+,s)+ D\partial_x\p_{\Psi}(0^-,s).
 \end{align}
 \end{subequations}
Recall that $\p_{\Psi}$ is related to the local time propagator according to equation (\ref{BCL}). Hence, for $x,x_0>0$,
\begin{align}
\partial_x\p_{\Psi}(x,s|x_0)&=\int_0^{\infty}\Psi(\ell) \partial_x \PP(x,\ell,s|x_0) d\ell\nonumber \\
&=\int_0^{\infty}\Psi(\ell)\left [\PP(x,0,s|x_0) \delta(\ell) +\frac{\partial \PP(x,\ell,s|x_0)}{\partial \ell}\right ]d\ell\nonumber \\
&=\int_0^{\infty}\psi(\ell)\PP(x,0,s|x_0) d\ell
\end{align}
with $\psi(\ell)=-\Psi'(\ell)$. We have used the boundary condition (\ref{oo}) and integration by parts. Substituting for $\PP$ using equation (\ref{Ptil}) gives
\begin{align}
\partial_x\p_{\Psi}(x,s|x_0)&=\frac{\widetilde{\psi}(0)}{2\sqrt{sD}}\left (\e^{-\sqrt{s/D}|x-x_0|}-\e^{-\sqrt{s/D}(x+x_0)}  \right )\nonumber \\
&\quad  +\frac{1}{D}\e^{-\sqrt{s/D}(x+x_0)}\widetilde{\psi}(\sqrt{s/D}).
\end{align}
Deriving the analogous equation for $x<0$ finally shows that
\begin{equation}
D\partial_x\p_{\Psi}(0^+,s)- D\partial_x\p_{\Psi}(0^-,s)=\widetilde{\psi}(\sqrt{s/D})\int_{-\infty}^{\infty}\e^{-\sqrt{s/D}|x_0|}f(x_0)dx_0=\Gamma_{\Psi}(s),
\end{equation}
since $\widetilde{\psi}(s)=1-s\widetilde{\Psi}(s)$. We deduce from equation (\ref{enc1}a) that $D \partial_x\widetilde{\rho}_{\psi}(0^+,s) = D\partial_x\widetilde{\rho}_{\Psi}(0^-,s)$. In other words, the flux through the membrane is continuous, as it is in the standard permeable boundary condition. Equation (\ref{enc1}b) then implies that
\begin{align}
D \partial_x\widetilde{\rho}_{\psi}(0^{\pm},s) &=\widetilde{\psi}(\sqrt{s/D})\left [\int_{0}^{\infty}\e^{-\sqrt{s/D}x_0}f(x_0)dx_0-\int_{-\infty}^0\e^{\sqrt{s/D}x_0}f(x_0)dx_0\right ]\nonumber \\
&=\frac{D \widetilde{\psi}(\sqrt{s/D})}{\widetilde{\Psi}(\sqrt{s/D})}[\p_{\Psi}(0^+,s)-\p_{\Psi}(0^-,s)]\nonumber \\
&=\frac{D \widetilde{\psi}(\sqrt{s/D})}{\widetilde{\Psi}(\sqrt{s/D})}[\widetilde{\rho}_{\Psi}(0^+,s)-\widetilde{\rho}_{\Psi}(0^-,s)].
\end{align}

\begin{figure}[t!]
  \centering
  \includegraphics[width=8cm]{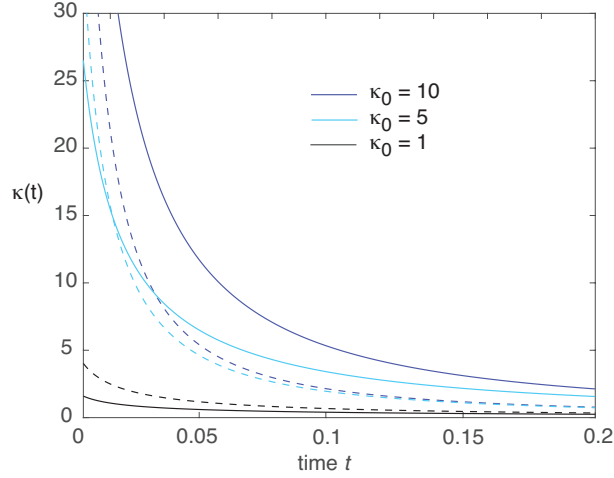}
  \caption{Plot of permeability function $\kappa(t)$ as a function of time $t$ for various values of $\kappa_0$ with $D=10$ (solid curves) and $D=1$ dashed curves.}
  \label{fig6}
\end{figure}
 
 In the exponential case, $\psi(\ell)=(\kappa_0/D)\Psi(\ell)$, we recover the permeable boundary condition (\ref{ser}). For non-exponential distributions, the boundary condition involves a time-dependent permeability. More specifically, setting
 \begin{equation}
 \label{Lkap}
 \widetilde{\kappa}(s)=\frac{D \widetilde{\psi}(\sqrt{s/D})}{\widetilde{\Psi}(\sqrt{s/D})}
 \end{equation}
 and using the convolution theorem, the boundary condition in the time domain takes the form
 \begin{align}
D \partial_x {\rho}_{\psi}(0^{\pm},t) 
&=\int_0^t \kappa(\tau)[ {\rho}_{\Psi}(0^+,t-\tau)- {\rho}_{\Psi}(0^-,t-\tau)]d\tau.
\end{align}
For the sake of illustration, suppose that $\psi(\ell) $ is given by the gamma distribution:
\begin{equation}
\label{psigam}
\psi(\ell)=\frac{\gamma(\gamma \ell)^{\mu-1}\e^{-\gamma \ell}}{\Gamma(\mu)}, \mu >0,
\end{equation}
where $\Gamma(\mu)$ is the gamma function. The corresponding Laplace transforms are
\begin{equation}
\widetilde{\psi} (z)=\left (\frac{\gamma}{\gamma+z}\right )^{\mu},\quad \widetilde{\Psi}(z)=\frac{1-\widetilde{\psi}(z)}{z}
\end{equation}
Here $\gamma$ determines the effective absorption rate. If $\mu=1$ then $\psi$ reduces to the exponential distribution with constant reactivity $\kappa_0 = D\gamma$. The parameter $\mu$ thus characterizes the deviation of $\psi(\ell)$ from the exponential case. If $\mu <1$ ($\mu>1$) then $\psi(\ell)$ decreases more rapidly (slowly) as a function of the local time $\ell$. Substituting the gamma distribution into equation (\ref{Lkap}) yields
 \begin{equation}
 \widetilde{\kappa}(s)=\frac{\sqrt{sD}\gamma^{\mu}}{(\gamma+\sqrt{s/D})^{\mu}-\gamma^{\mu}}.
 \end{equation}
 If $\mu=1$ then $\widetilde{\kappa}(s)=\gamma D=\kappa_0$ and $\kappa(\tau)=\kappa_0\delta(\tau)$. An example of $\mu\neq 1$ that has a simple inverse Laplace transform is $\mu=2$:
  \begin{equation}
 \widetilde{\kappa}(s)=\frac{D\sqrt{D}\gamma^2}{2\sqrt{D}\gamma+\sqrt{s}}=\frac{\kappa_0^2/\sqrt{D}}{2\kappa_0/\sqrt{D}+\sqrt{s}}
 \end{equation}
 and
 \begin{align}
 \kappa(\tau)=\frac{\kappa_0^2}{\sqrt{D}}\left [\frac{1}{\sqrt{\pi \tau}}-\frac{2\kappa_0}{\sqrt{D}}\e^{4\kappa_0^2\tau/D} \mbox{erfc}(2\kappa_0\sqrt{\tau/D})\right ],
 \end{align}
 where $\mbox{erfc}(x)=(2/\sqrt{\pi})\int_x^{\infty} \e^{-y^2}dy$ is the complementary error function. Example plots of $\kappa(\tau)$ are shown in Fig. \ref{fig6}. It can be seen that $\kappa$ is an exponentially decaying function of time whose rate of decay depends on $\kappa_0$ and $D$.

\setcounter{equation}{0} 
 \section{Conclusion} In this paper we have developed a general probabilistic framework for modeling one-dimensional diffusion through semi-permeable membranes. We took as our starting point the snapping out BM recently introduced by Lejay \cite{Lejay16}. The latter sews together successive rounds of partially reflecting BM in either the positive or negative $x$ domains. The major advantage of this formulation is that the probability density of particle position satisfies a renewal equation that can be generalized by appropriate modifications of the underlying partially reflected BM. As our first example, we considered partially reflected BM with stochastic resetting, which resulted in a diffusion process through a semipermeable membrane with a nontrivial NESS. Although the NESS was independent of the permeability $\kappa_0$, the associated relaxation process was $\kappa_0$-dependent. Our second example used an encounter-based method to modify the absorption process that kills a given round of partially reflected BM. This resulted in diffusion through a semipermeable membrane with a time-dependent permeability.

\begin{figure}[b!]
  \centering
  \includegraphics[width=5cm]{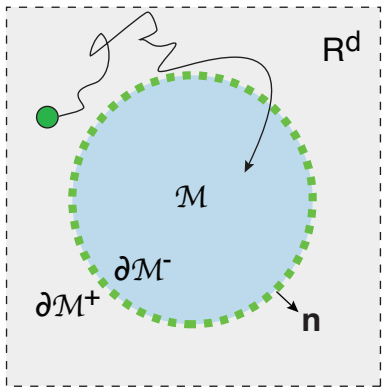}
  \caption{Example configuration for diffusion through a higher-dimensional semipermeable membrane.}
  \label{fig7}
\end{figure}

Although we focused on one-dimensional diffusion processes, the basic renewal equation framework  generalizes to higher spatial dimensions. However, the analysis is significantly more difficult. (Indeed most studies of skew BM and its generalizations are based on one-dimensional diffusions. A discussion of some mathematical papers on higher-dimensional skew BM can be found in \cite{Lejay16}.)  For the sake of illustration, consider diffusion in $\R^d$ that contains a closed bounded subdomain $\calM$. We treat the boundary $\partial \calM$ separating the two open domains $\R^d\backslash \overline{\calM}$ and $\calM$ as a semipermeable membrane with $\partial \calM^+ $ ($\partial \calM^-$) denoting the side approached from outside (inside) $\calM$. The higher-dimensional version of equation (\ref{class}) is then
\begin{subequations}
\label{dclass}
\begin{align}
\frac{\partial u(\x,t)}{\partial t}&=D\nabla^2u(\x,t) \quad \x\in {\mathbb G}\equiv (\R^d\backslash \overline{\calM})\cup \calM,\\
J(\x^{\pm},t)&=\kappa_0[u(\x^-,t)-u(\x^+,t)],\quad \x^{\pm} \in \partial \Omega^{\pm},
\end{align}
\end{subequations}
where $J(\x,t)=-D\nabla u(\x,t) \cdot \n$ and $\n$ is the unit normal directed out of $\calM$, see Fig. \ref{fig7}.

The major difference from the one-dimensional case is that it is now necessary to keep track of where on the boundary each round of partially reflected BM is killed, and from where the next round is initiated. In particular, suppose that whenever partially reflected BM is killed at a point $\y^+ \in \partial \calM^+$, a new round is immediately started from either $\y^+$ or $\y^-$ with equal probability etc. The higher-dimensional version of the last renewal equation (\ref{renewal}) is then
\begin{subequations}
\label{drenewal}
\begin{align}
 \rho(\x,t)&=p(\x,t)+\frac{\kappa_0}{2}\int_0^t \left \{ \int_{\partial \calM}p(\x,\tau|\y^+)[\rho(\y^+,t-\tau ) +\rho(\y^-,t-\tau )]d\y\right \}d\tau ,\nonumber \\
 &\quad  \x\in \R^d\backslash \overline{\calM} ,\\
 \rho(\x,t)&=q(\x,t)+\frac{\kappa_0}{2}\int_0^t \left \{ \int_{\partial \calM}q(\x,\tau|\y^-)[\rho(\y^+,t-\tau ) +\rho(\y^-,t-\tau )]d\y\right \}d\tau ,\nonumber \\
 &\quad  \x\in  \calM ,
   \end{align}
 \end{subequations}
 where $p(\x,t|\y)$ and $q(\x,t|\y)$ are the probability densities for partially reflected BM in the domains $R^d\backslash \overline{\calM}$ and $\calM$, respectively. In addition
 \begin{equation}
 p(\x,t)=\int_{R^d\backslash \overline{\calM}}p(\x,t|\x_0)g(\x_0)d\x_0,\ q(\x,t)=\int_{ \calM}q(\x,t|\x_0)g(\x_0)d\x_0.
 \end{equation}
 where $g(\x_0)$ is the initial probability density in ${\mathbb G}$. Elsewhere we will show that the solution $\rho(\x,t)$ of the integral equation (\ref{drenewal}) satisfies a BVP of the form (\ref{dclass}). This will allow us to introduce stochastic resetting and encounter-based models of absorption in an analogous fashion to the 1D case. However, finding an explicit solution for $\rho$ is more difficult than the 1D case, even after Laplace transforming.
 One exception is taking $\partial \calM$ to be a $(d-1)$-dimensional sphere and using spherical symmetry. This recovers a renewal equation similar in form to (\ref{renewal}) with $x$ replaced by the radial coordinate. Another possibility is to Laplace transform the renewal equation and carry out a Neumann series expansion of the integral equation in $\y$ for small $\kappa_0$.


\begin{thebibliography}{9}


\bibitem{Alberts15} Alberts B, Johnson A, Lewis J, Morgan D, Raff M, Roberts K, Walter P. 2015 {\em Molecular biology of the cell. 6th ed.} Chapter 11. New York: Garland Science

\bibitem{Philips12} Phillips R, Kondev J, Theriot J, Garcia HG, Orme N. 2012 
{\em Physical Biology of the Cell} Garland, New York 

\bibitem{Bressloff21} Bressloff PC 2021 {\em Stochastic Processes in Cell Biology}. Springer Switzerland

\bibitem{Kusumi05} Kusumi A, Nakada C, Ritchie K, Murase K, Suzuki K,
Murakoshi H, Kasai RS, Kondo J, Fujiwara T. 2005 
Paradigm shift of the plasma membrane concept from the two-dimensional
continuum fluid to the partitioned fluid: high-speed
single-molecule tracking of membrane molecules {\em Annu. Rev.
Biophys. Biomol. Struct. } {\bf 34} 351 

\bibitem{Evans02} Evans WJ, Martin PE. 2002 {Gap junctions: structure and function.} {\em Mol. Membr. Biol.} {\bf 19} 121-136 

\bibitem{Connors04} Connors BW, Long MA 2004 {Electrical synapses in the mammalian brain.} {\em Ann. Re. Neurosci.} {\bf 27} 393-418  

\bibitem{Good09} Goodenough DA, Paul DL. 2009 {Gap junctions.} Cold Spring Harb Perspect Biol {\bf 1} a002576  





\bibitem{Beyer16} Beyer HL, Gurarie E, B\o rger L, Panzacchi M, Basille M,
Herfindal I, Van Moorter B, Lele SR, Matthiopoulos J 2016
``You shall not pass!'': quantifying barrier permeability and proximity
avoidance by animals {\em J. Anim. Ecol.} {\bf 85} 43  

\bibitem{Assis19} Assis JC, Giacomini HC, Ribeiro MC. 2019 Road permeability
index: evaluating the heterogeneous permeability of
roads for wildlife crossing {\em Ecol. Indic.} {\bf 99} 365

\bibitem{Kenkre21} Kenkre VM, Giuggioli L. 2021 {\em Theory of the Spread of Epidemics
and Movement Ecology of Animals: An Interdisciplinary
Approach Using Methodologies of Physics and Mathematics}
Cambridge University Press, Cambridge, UK.

\bibitem{Tanner78} Tanner JE. 1978 Transient diffusion in a system partitioned
by permeable barriers. application to nmr measurements
with a pulsed field gradient. {\em J. Chem. Phys.} {\bf 69} 1748


\bibitem{Brink85} Brink PR, Ramanan SV. 1985 A model for the diffusion of fluorescent probes in the septate giant axon of earthworm: axoplasmic diffusion and junctional membrane permeability. {\em Biophys. J.} {\bf 48} 299-309  

\bibitem{Ramanan90}  Ramanan SV, Brink PR. 1990 Exact solution of a model of diffusion in an infinite chain or monlolayer of cells coupled by gap junctions. {\em Biophys. J.} {\bf 58} 631-639  




\bibitem{Kos01} Kosztolowicz T, Mrowczynski S 2001 Membrane boundary
condition {\em Acta Physica Polonica. Series B} {\bf 32} 217  

\bibitem{Moutal19} Moutal N, Grebenkov, DS 2019 Diffusion across semi-permeable
barriers: spectral properties, efficient computation, and applications,
{\em J. Sci. Comput.} {\bf 81} 1630.

\bibitem{Powles92} Powles JG, Mallett M, Rickayzen G, Evans W. 1992 Exact
analytic solutions for diffusion impeded by an infinite array of
partially permeable barriers {\em Proc. R. Soc. Lond. A} {\bf 436} 391


\bibitem{Kenkre08} Kenkre VM, Giuggioli L, Kalay Z. 2008 Molecular motion
in cell membranes: analytic study of fence-hindered random
walks {\em Phys. Rev. E} {\bf 77} 051907 

\bibitem{Kay22} Kay T, Giuggioli 2022 Diffusion through permeable interfaces: Fundamental equations and their application to
first-passage and local time statistics. {\em Phys. Rev. Res.} {\bf 4} L032039




   \bibitem{Ito65} Ito K, McKean HP 1965 {\em Diffusion Processes and Their Sample Paths} Springer-Verlag,
Berlin
  
   \bibitem{Freidlin85} Freidlin M. 1985 {\em Functional Integration and Partial Differential Equations}
Annals of Mathematics Studies, Princeton University Press, Princeton
New Jersey

\bibitem{Papanicolaou90} Papanicolaou VG. 1990 {The probabilistic solution of the third boundary
value problem for second order elliptic equations} {\em Probab. Th. Rel. Fields}
{\bf 87}, 27-77 

 \bibitem{Milshtein95} Milshtein GN. 1995 {The solving of boundary value problems by numerical
integration of stochastic equations.} {\em Math. Comp. Sim.} {\bf 38} 77-85


\bibitem{Borodin96} Borodin AN, Salminen P. 1996 {\em Handbook of Brownian Motion: Facts and Formulae}
Birkhauser Verlag, Basel-Boston-Berlin.

\bibitem{Grebenkov06} Grebenkov DS 2006 Partially Reflected Brownian Motion: A Stochastic
Approach to Transport Phenomena, in ``Focus on Probability Theory'',
Ed. Velle LR pp. 135-169 (Hauppauge: Nova Science Publishers)



\bibitem{Lejay16} Lejay A. 2016 The snapping out Brownian motion. {\em The Annals of Applied Probability}
 {\bf 26} 1727-1742.
 
 \bibitem{Ito63} Ito K, McKean H. 1963 Brownian motions on a half line {\em Illinois J.Math.} {\bf 7} 181-231
 
 \bibitem{Lejay06} Lejay A. 2016 On the constructions of the skew Brownian motion {\em Probab. Surv.} {\bf 3} 413-466.
 
\bibitem{Decamps06} Decamps M, Goovaerts M, Schoutens W. 2006 Asymmetric skew Bessel processes and their applications to finance, {\em J. Comput. Appl. Math.} {\bf 186} 130-147.
 
\bibitem{App11} Appuhamillage T, Bokil V, Thomann E, Waymire E, Wood B. 2011 Occupation and local times for skew Brownian motion with applications to dispersion across an interface. {\em  Ann. Appl. Probab.} {\bf 21} 183-214.

\bibitem{Gairat17} Gairat A, Shcherbakov V. 2017 Density of skew Brownian motion and its functionals with application in finance. {\em Math. Finance} {\bf 27} 1069-1088

\bibitem{Evans20} Evans, M. R., Majumdar, S. N., Schehr, G.: Stochastic resetting and applications. {\em J. Phys. A: Math. Theor.} {\bf 53} 193001 (2020).






\bibitem{Grebenkov20} Grebenkov DS. 2020  {Paradigm shift in diffusion-mediated surface phenomena.} {\em Phys. Rev. Lett.} {\bf 125}, 078102  


\bibitem{Grebenkov22} Grebenkov DS. 2022  {An encounter-based approach for restricted diffusion with a gradient drift.}  {\em J. Phys. A.} {\bf 55} 045203 

\bibitem{Bressloff22} Bressloff PC. 2022  Diffusion-mediated absorption by partially reactive targets: Brownian functionals and generalized propagators. {\em J. Phys. A.} {\bf 55} 205001

\bibitem{Bressloff22a} Bressloff PC 2022 Spectral theory of diffusion in partially absorbing media. {\em Proc. R. Soc. A} {\bf 478} 20220319


\bibitem{Evans13} Whitehouse J, Evans M R and Majumdar SN 2013. Effect of partial absorption on diffusion with resetting {\em Phys. Rev. E} {\bf 87} 022118. 

\bibitem{Bressloff22b} Bressloff PC 2022 Diffusion-mediated surface reactions and stochastic resetting. {\em J. Phys. A} {\bf 55} 275002 


 \bibitem{Evans11a} Evans M R and Majumdar S N 2011 Diffusion with stochastic resetting {\em Phys. Rev. Lett.}{\bf 106} 160601.

\bibitem{Evans11b} Evans M R and Majumdar S N 2011 Diffusion with optimal resetting {\em J. Phys. A Math. Theor.} {\bf 44} 435001. 




\bibitem{Roldan17} Roldan E, Gupta S. 2017 Path-integral formalism for stochastic resetting: Exactly solved
examples and shortcuts to confinement {\em Phys. Rev. E} {\bf 96} 022130

\bibitem{Pinsky20}  Pinsky RG 2020 Diffusive search with spatially dependent resetting. {\em Stochastic Processes and
their Applications} {\bf 130} 2954-2973

\bibitem{Majumdar15} Majumdar S N, Sabhapandit S, Schehr G. 2015 Dynamical transition in the temporal relaxation of stochastic processes under resetting. {\em Phys. Rev. E} {\bf 91} 052131 




\bibitem{Bressloff21a} Bressloff PC. 2021 Accumulation time of stochastic processes with resetting. {\em J. Phys. A} {\bf 54} 354001  

\bibitem{Berez10} Berezhkovskii A M, Sample C and Shvartsman S Y 2010 How
long does it take to establish a morphogen gradient?
{\em Biophys. J.} {\bf 99} L59-L61 



\bibitem{Berez11} Berezhkovskii A M, Sample C and Shvartsman S 2011 Formation
of morphogen gradients: local accumulation time.
{\em Phys Rev E} {\bf 83} 051906 

\bibitem{Gordon11} Gordon P, Sample C, Berezhkovskii A M, Muratov C B and Shvartsman S 2011 Local kinetics of morphogen gradients.
{\em Proc Natl Acad Sci.}  {\bf 108} 6157-6162  




\end{thebibliography}
\end{document}